\documentclass[sigconf, 10pt, screen]{acmart} 

\usepackage{tikz}
\usepackage[utf8]{inputenc}
\usepackage{url}
\usepackage{caption}
\usepackage{cleveref}
\usepackage{graphicx}
\usepackage{subfig}
\usepackage{float}
\usepackage{balance}
\usepackage{enumitem}
\usepackage{bm}

\newcommand{\sandbox}{FedApp} 

\AtBeginDocument{%
  \providecommand\BibTeX{{%
    \normalfont B\kern-0.5em{\scshape i\kern-0.25em b}\kern-0.8em\TeX}}}

\renewcommand\footnotetextcopyrightpermission[1]{} 
\setcopyright{none}

\begin{document}

\pagestyle{plain}

\title{\sandbox{}: a Research Sandbox for Application Orchestration in Federated Clouds using OpenStack}

\author{Johan Ruuskanen}
\affiliation{%
  \institution{Department of Automatic Control, Lund University}
  \city{Lund}
  \country{Sweden}
}
\email{johan.ruuskanen@control.lth.se}

\author{Haorui Peng}
\affiliation{%
  \institution{Department of Electrical and Information Technology, \\Lund University}
  \city{Lund}
  \country{Sweden}
}
\email{haorui.peng@eit.lth.se}

\author{Alfred Åkesson}
\affiliation{%
 \institution{Department of Computer Science, Lund University}
  \city{Lund}
  \country{Sweden}
 }
 \email{alfred.akesson@cs.lth.se}

\author{Lars Larsson}
\affiliation{%
  \institution{Department of Computing Science, Ume\aa~University}
  \city{Ume\aa}
  \country{Sweden}
 }
 \email{larsson@cs.umu.se}

\author{Maria Kihl}
\affiliation{%
  \institution{Department of Electrical and Information Technology,\\ Lund University}
  \city{Lund}
  \country{Sweden}
}
\email{maria.kihl@eit.lth.se}

\begin{abstract}
    Multi-cluster federation is envisioned to be the next-generation cloud infrastructure, where it will play a vital part in the realization of concepts such as edge and fog computing. Orchestrating applications in federated environments poses new challenges to well-known research problems in various fields, such as load-balancing, auto-scaling, resource allocation and service migration. However, as access to real multi-cluster infrastructure is limited, a test-bed that provides similar characteristics to a real system is in demand. To enable researchers in associated fields to quickly setup experiments in a federated cloud environment, we have created the open-source sandbox \sandbox{} that simplifies the process of deploying multiple virtual clusters in an OpenStack environment with the possibility of adding realistic network characteristics between sites. Each cluster comes deployed with the open-source and production-grade container orchestrator Kubernetes, complete with federation-wide monitoring using Prometheus/Grafana and simplified inter-cluster microservice communication using Istio.
\end{abstract}

\keywords{Multi-cluster Federation, Edge Computing, Orchestration, Kubernetes, Sandbox}

\maketitle

\section{Introduction}

It has long been known that relying on a single cloud provider is not feasible for certain applications and use-cases~\cite{rochwerger2008reservoir,ferrer2012optimis}. Instead, multiple clusters can be used in a joint effort to offer increased performance. This collaboration between a group of autonomous clusters of computational resources is commonly referred to as cluster \emph{federation}. Together with emerging infrastructure provider alternatives such as edge computing~\cite{shi2016edge,villari2016osmotic}, driven in part by the 5G evolution~\cite{hu2015mobile}, it is increasingly more important to make separate clouds cooperate to jointly provide applications in a more desirable manner. Edge locations are further destined to outnumber the low number of major public cloud providers and regions, thus causing a more pressing need for software to manage underlying systems~\cite{cherrueau2018edge}. To put things into perspective, as of June 2020 Google offers cloud infrastructure in a mere 23 regions. In contrast, edge computing opens the Infrastructure-as-a-Service market to Telco providers, and thus dramatically changes the scale and nature of the infrastructure being offered to customers. In Germany alone, a single Telco provider (Deutsche Telekom) will operate 36,000 base stations by 2021.~\cite{kiel-koslowski_deutsche_2019}. Regardless of whether all or ``merely'' ten percent of these will offer edge computing facilities, it is clear that methodologies and technologies supporting management and use of such infrastructure will require novel research. 

Regarding these new federated environments together with recent trends in shifting application architecture from single monolithic implementations to collections of networked microservices, which in itself adds new complexities on cluster management \cite{Fazio2016, Gan2019}, new questions arise in how to automatically orchestrate applications to provide services of desirable performances. As cluster federations are by nature highly dynamic environments, where uncertain cluster-to-cluster connections with potentially non-stationary latency on top of existing intra-cluster dynamics, application robustness will be an issue. Strategies for inter-cluster load-balancing and routing between services, or auto-scaling of individual services will need to adapt to changing dynamics in real-time to guarantee performance. Further, new problems arise in service scheduling and migration when an entire federation of clusters is considered. Objectives and constraints in end-to-end latency and network bottlenecks might exist and be sporadically violated due to the dynamics. 

These types of research problems are commonly tackled using simulation or emulation models~\cite{Ahmed14, PerezAbreu2019, ramprasad2019emuiot}, which are often quick to set up and play an important part for initial algorithm design and evaluation, but ultimately fail to capture actual system behaviour. To complement this, researchers have the option to perform actual experiments in real environments. However, creating such an environment is in general difficult, expensive and time-consuming, and the steps required to go beyond simulations to explore ideas in real settings quickly becomes complicated when dealing with complex systems such as multi-cluster federation. To the best of our knowledge, there is currently no easy way for academic researchers to cross this gap and access a desired federation of clusters. 

An interesting research question then arises in how a research prototype for federated cloud environments can be designed to best support application orchestration research. To this end, we have created the sandbox \sandbox{} with the goal of providing such an environment that (a) is flexible, yet remains a close approximation of real systems, (b) is easy to both deploy and use, and (c) has an easily extendable implementation. Its key features include:

\begin{enumerate}
    \item creation of a user-defined, multi-cluster virtual environment in OpenStack, with a centralized structure complete with tools for controlling and monitoring the entire federation, streamlining the setup and usage of the sandbox.
    \item possibility of inducing network characteristics such as delay, jitter and packet loss between clusters, thus enabling faithful emulation of real-world federations.
    \item application orchestration using Kubernetes complete with a multi-cluster service mesh layer using Istio, for easy handling and deployment of federation-wide applications.
\end{enumerate}
\textbf{The sandbox can be found here\footnote{\url{https://github.com/JohanRuuskanen/FedApp}}}.

The rest of the paper is organized as follows; In Section \ref{sec:design} we go into detail on how the sandbox is designed and how it relates to our goal. In Section \ref{sec:researchfunc} we then present how the sandbox can enable research in federated application orchestration. Further, in Section \ref{sec:proofofconcept} we present a proof-of-concept fog computing scenario to showcase the sandbox. 

\section{Sandbox design} \label{sec:design}

\begin{figure}[t!]
	\centering
    \includegraphics[width=0.8\linewidth]{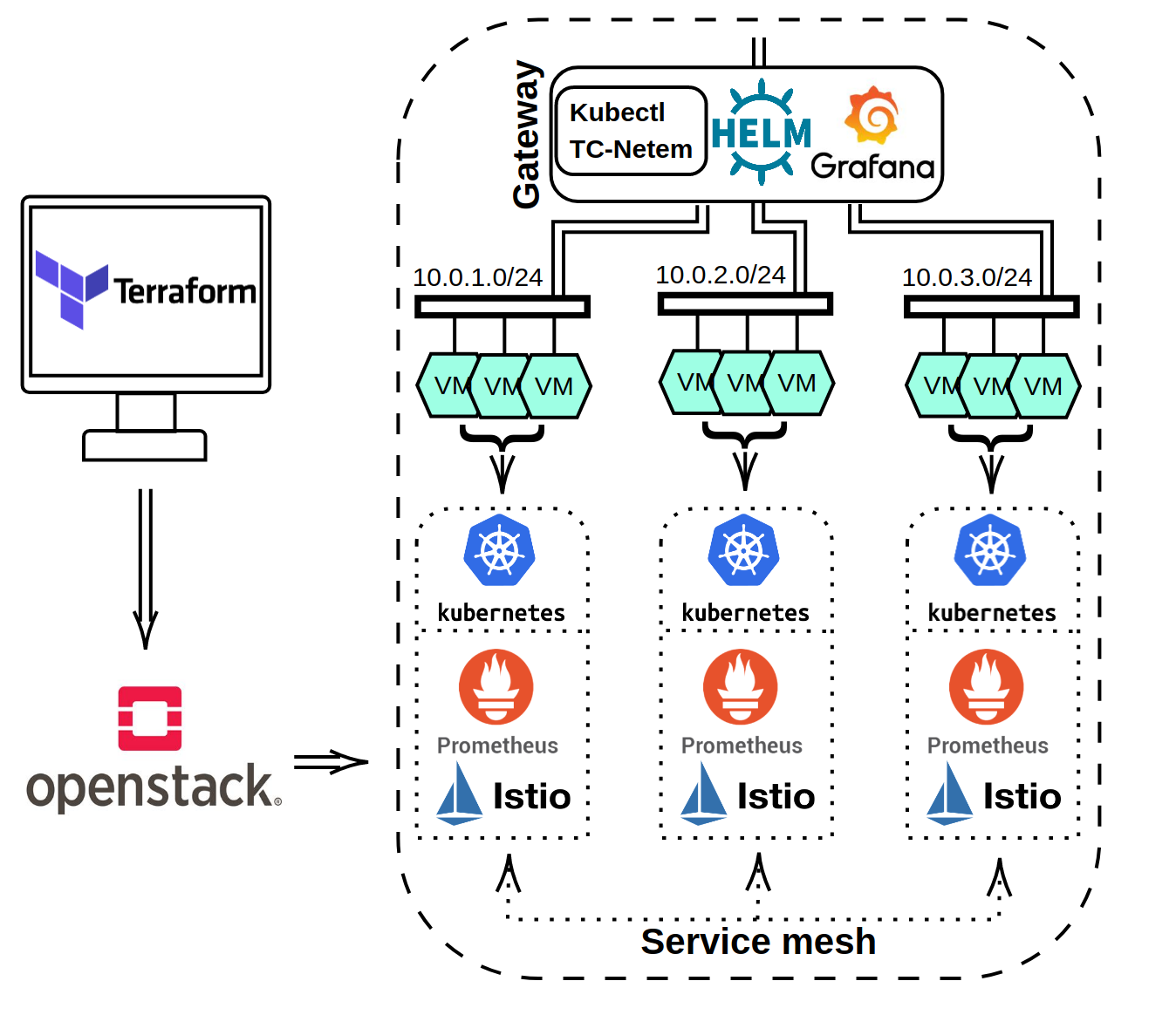}
	\caption{Illustration of the complete sandbox with 3 clusters containing 3 virtual machines each.}
	\label{fig:sandbox_setup}
	   \noindent\makebox[\linewidth]{\rule{\linewidth}{0.4pt}}
\end{figure}

In this section, the design choices for creating the federation environment to fulfill our goal is presented. We first discuss how our emulated federation environment is constructed, later we explain how we can induce realistic network characteristics, and finally we describe how the sandbox simplifies the deployment and management of federation-wide applications. An illustration of the complete setup can be seen in Figure \ref{fig:sandbox_setup} which can be used as an overview of the sandbox in its entirety. 

\subsection{Topology of the emulated federation} \label{sec:topology}

The sandbox will need some type of infrastructure to be deployed upon. To provide this in a both flexible and easy to use manner, we chose to supply the means of creating a suitable virtual infrastructure in OpenStack. OpenStack is an open-source software platform for providing common IaaS (Infrastructure-as-a-Service) on physical hardware, and is commonly used to operate private clouds. This platform provides both the means of creating a controlled environment in which the rest of the sandbox can be deployed in a standardized manned, and makes it flexible as there is a standardized way of scaling the size and amount of clusters. 
To emulate a federated cloud setup, each cluster is first represented and provisioned in OpenStack as a set of virtual machines (VM) with their own isolated internal network. To further provide a close approximation to a real cluster federation, it is vital that users are able to impose desired network characteristics between clusters. To enable this we let each internal cluster network be connected to a gateway virtual machine, which allows us to use it as a single point handler of routing and network emulation between clusters. 

Given the stated goal of this project, centralizing the network through a VM in this manner is beneficial. First, it enables researchers to affect network characteristics and observe network behaviour such as traffic patterns and workload classifications from a single point. 
Moreover, a centralized VM increases usability as it can be shipped with all the tools necessary for controlling, monitoring and deploying applications on the clusters. Finally, the choice of pooling all functionalities for commandeering the emulated federation makes the setup easy to extend, and easy to debug when something eventually breaks down. 

\subsection{Network characteristics emulation} \label{sec:netem}

Benefiting from the chosen centralized topology, the procedure to mimic a real-world networking environment is greatly simplified. Each cluster sees itself as a ``stand-alone'' cluster on a private network, where all the inter-cluster communication traffic is handled by the Gateway VM. Thus by only affecting the inter-cluster routing logic on the gateway, an arbitrary inter-cluster network profile could be emulated. 

We achieve this by fully utilizing the Linux Traffic Control (TC) utility on the gateway. The TC network emulator (TC-netem)\footnote{\url{https://www.linux.org/docs/man8/tc-netem.html}} provides the possibility of adding packet loss, delay and other characteristics on the packets from a selected Network Interface Controller (NIC). By applying desired emulated characteristics on the NICs of the Gateway VM, the clusters are not aware of the network status in beforehand. Thereby, the network characteristic to a cluster can be specified arbitrarily, to for example mimic the propagation and transmission delay caused by the geographic distance. Further, by deploying TC classful queuing discipline (qdisc) and filter, it is possible to control the point-to-point network characteristics between clusters. Each class can then be given its individual configuration for a desired cluster-to-cluster network characteristic without affecting important meta-communication between the Gateway VM and the clusters, such as monitoring data or control commands. 

Using TC-netem, it is then possible to both emulate real-world network characteristics, and to test some customized scenarios, by applying delay and loss along with their distributions to the network connections between any desired pair of clusters.

\subsection{Managing federation-wide applications} \label{sec:implementation_details}

Cloud applications are increasingly being packaged and deployed as collections of networked services using containerization technologies. To allow such applications to be efficiently managed, the sandbox fits each cluster with the well-known container orchestrator Kubernetes, which creates an abstraction layer between the raw infrastructure and applications. This provides both a vital platform for building, deploying and handling federation-wide applications in a standardized manner, and allows researchers to experiment in a popular and well-maintained open-source setting. To enable non-trivial collaboration in service-graphs spanning multiple clusters, the sandbox deploys Istio on top of Kubernetes, which greatly simplifies handling inter-cluster service-to-service networking. The sandbox further supplies kubectl at the Gateway VM, enabling full control of the Kubernetes clusters via a CLI. Finally, to simplify the management of more advanced deployments, the Gateway VM is supplied with the Kubernetes package manager Helm. 

To provide insight on the behaviour of the emulated federation and its deployed applications, the sandbox supplies federation-wide monitoring in the form of data-collection using Prometheus, and a Grafana GUI deployed on the Gateway VM. With this setup, it is possible to in real-time inspect important metrics such as CPU and memory utilization out-of-the-box for all parts of the federation and its applications from the Gateway VM. Further, the Istio service mesh makes it possible to retrieve the timestamp of every request entering/exiting any microservice without tampering with the code of the service. It is also possible to collect tracing information of all requests, by performing the minimal code change of forwarding http headers in each service. 

As we set out to make the sandbox flexible, easy to use and extendable, its deployment is heavily scripted and requires little input from the user to create the multiple Kubernetes clusters with Istio and monitoring as showcased in \Cref{fig:sandbox_setup}. To setup, the only things needed are access to an OpenStack cluster and a local computer with Terraform and Ansible installed. The deployment itself further capitalizes on the natural modularity between the different tools, which makes it easy to change or debug one part of the deployment without affecting the other parts of the sandbox.

\section{Functionalities}
\label{sec:researchfunc}
The sandbox was created with the goal of providing a federated cloud environment for application orchestration research. In this section we discuss its functionalities that could specifically contribute to this point. Via our design, researchers can set up an user-defined federated Kubernetes environment, with a single centralized point for collecting all tools and scripts needed for researchers to deploy, monitor and control federation-wide applications.

The decision to provision the infrastructure with OpenStack together with TC-netem for network emulation enables researchers to experiment in a wide range of federated cloud or fog computing settings. Further, the usage of Kubernetes and Istio makes it possible to easily deploy and control federation-wide applications in these settings. Finally, together with the Istio service mesh, the inclusion of Prometheus makes it possible to effortlessly monitor both the resource utilization for every service and VM, and further also the timestamp and tracing for every service. Because of the centralized structure and our emphasise on extendability, new tools or software that are not supplied out-of-the-box can easily be incorporated into the sandbox.

The possible use cases of \sandbox{} is numerous, and many of the open issues in cloud computing as discussed in \cite{Buyya18} could readily be explored in it. Below follows some key research problems that can directly take advantage of experimental investigation and validation in \sandbox.

\begin{enumerate}
    \item \textbf{Resource management:} Auto-scaling and load-\\balancing in large scale services and multi-cluster settings is an open issue. As both Kubernetes and Istio natively supports these methods, the sandbox gives a great entrypoint for researchers to test their own solutions. 
    
    \item \textbf{Scheduling:} Where to schedule what microservice in a service graph, the effects of multi-tenancy and over-provisioning, and how to tackle migration problems in a dynamic multi-cloud setting can readily be investigated using the available control and monitoring solutions in the sandbox. 

    \item \textbf{Reliability:} The sandbox provides a natural platform for developing detection and recovering solutions for single-point fail-over in federated cloud environments. For example, TC-netem can be used to impose a connection drop to a cluster without affecting the meta-communcation between the Gateway VM. Further, the fine grained tracing information makes it possible to evaluate performance models of e.g. the quality-of-service in applications.
    
    \item \textbf{Networking:} Analysis of network traffic in large scale distributed systems can readily be performed at the Gateway VM, as it handles all cluster-to-cluster communication. Further, TC-netem offers the opportunity to customize various network characterises to evaluate systems and applications under different types of networking scenarios. 

    \item \textbf{Security:} The sandbox supports the study and analysis of security and privacy on cloud applications in a federated setting. For example, Istio provides a natural way of imposing encryption on service-to-service communication. 

\end{enumerate}

\section{Proof-of-concept}
\label{sec:proofofconcept}

To showcase some of the functionalities and to demonstrate how the sandbox can be used, we in this section present a fog computing scenario with a federated example application deployed on top of the sandbox. In the scenario we will consider a face detection application that in order to save bandwidth, has had its image pre-processing off-loaded to two smaller edge clusters $C_1, C_2$ from the actual classification at two larger data centers $C_3, C_4$.

\subsection{Deploying the scenario}

Without loss of generality, each virtual edge cluster and data center were given the same size of four VMs each with four vCPUs and 8 Gbytes of RAM. The Gateway VM was further deployed with four vCPUs and 16 Gbytes of RAM. All of the VMs were given Ubuntu 18.04 as the instance image. In order to emulate a mock geographic distance, a delay was introduced using TC-netem between the edge clusters $C_1, C_2$ and the backend cluster $C_3$ with the following one-way delay matrix. 
\begin{align*}
    D = \begin{pmatrix}
        0 & 0 & 25 & 0 \\
        0 & 0 & 25 & 0 \\
        25 & 25 & 0 & 0 \\
        0 & 0 & 0 & 0
    \end{pmatrix}~ms
\end{align*}
where $D_{ij}$ gives the one way delay between clusters $C_i$ and $C_j$. This means that the roundtrip-time (RTT) between $C_1, C_2$ and $C_3$ should be 50~ms. 

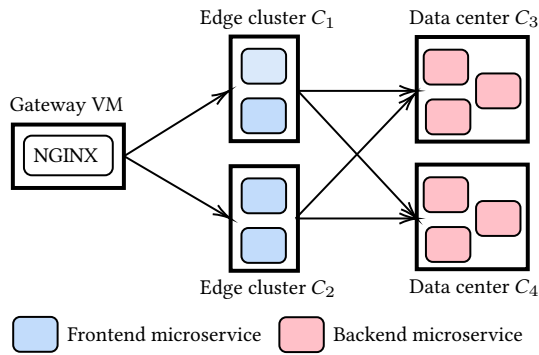
\begin{figure}[!t]
    \centering
    \scalebox{1.0}{
    \tikzset{every picture/.style={line width=0.75pt}} 

\begin{tikzpicture}[x=0.75pt,y=0.75pt,yscale=-0.8,xscale=0.8]
\footnotesize

\draw [line width=1.5pt]  (201,28.71) -- (243.43,28.71) -- (243.43,94.71) -- (201,94.71) -- cycle ;
\draw  [fill={rgb, 255:red, 182; green, 211; blue, 245 }  ,fill opacity=0.5]   (207.97,40.83) .. controls (207.97,38.4) and (209.94,36.43) .. (212.37,36.43) -- (231.57,36.43) .. controls (234,36.43) and (235.97,38.4) .. (235.97,40.83) -- (235.97,54.03) .. controls (235.97,56.46) and (234,58.43) .. (231.57,58.43) -- (212.37,58.43) .. controls (209.94,58.43) and (207.97,56.46) .. (207.97,54.03) -- cycle ;
\draw [fill={rgb, 255:red, 133; green, 186; blue, 248 }  ,fill opacity=0.5 ]  (207.83,72.11) .. controls (207.83,69.68) and (209.8,67.71) .. (212.23,67.71) -- (231.43,67.71) .. controls (233.86,67.71) and (235.83,69.68) .. (235.83,72.11) -- (235.83,85.31) .. controls (235.83,87.74) and (233.86,89.71) .. (231.43,89.71) -- (212.23,89.71) .. controls (209.8,89.71) and (207.83,87.74) .. (207.83,85.31) -- cycle ;

\draw [line width=1.5pt]  (318.33,29) -- (388.33,29) -- (388.33,95) -- (318.33,95) -- cycle ;
\draw [fill={rgb, 255:red, 255; green, 129; blue, 145 }  ,fill opacity=0.5]  (322.73,41.4) .. controls (322.73,38.97) and (324.7,37) .. (327.13,37) -- (346.33,37) .. controls (348.76,37) and (350.73,38.97) .. (350.73,41.4) -- (350.73,54.6) .. controls (350.73,57.03) and (348.76,59) .. (346.33,59) -- (327.13,59) .. controls (324.7,59) and (322.73,57.03) .. (322.73,54.6) -- cycle ;
\draw [fill={rgb, 255:red, 255; green, 129; blue, 145 }  ,fill opacity=0.5]  (355.73,56.4) .. controls (355.73,53.97) and (357.7,52) .. (360.13,52) -- (379.33,52) .. controls (381.76,52) and (383.73,53.97) .. (383.73,56.4) -- (383.73,69.6) .. controls (383.73,72.03) and (381.76,74) .. (379.33,74) -- (360.13,74) .. controls (357.7,74) and (355.73,72.03) .. (355.73,69.6) -- cycle ;
\draw [fill={rgb, 255:red, 255; green, 129; blue, 145 }  ,fill opacity=0.5]  (324.13,72.8) .. controls (324.13,70.37) and (326.1,68.4) .. (328.53,68.4) -- (347.73,68.4) .. controls (350.16,68.4) and (352.13,70.37) .. (352.13,72.8) -- (352.13,86) .. controls (352.13,88.43) and (350.16,90.4) .. (347.73,90.4) -- (328.53,90.4) .. controls (326.1,90.4) and (324.13,88.43) .. (324.13,86) -- cycle ;
\draw  [line width=1.5pt] (318.33,109.5) -- (388.33,109.5) -- (388.33,175.5) -- (318.33,175.5) -- cycle ;
\draw [fill={rgb, 255:red, 255; green, 129; blue, 145 }  ,fill opacity=0.5]  (322.73,121.9) .. controls (322.73,119.47) and (324.7,117.5) .. (327.13,117.5) -- (346.33,117.5) .. controls (348.76,117.5) and (350.73,119.47) .. (350.73,121.9) -- (350.73,135.1) .. controls (350.73,137.53) and (348.76,139.5) .. (346.33,139.5) -- (327.13,139.5) .. controls (324.7,139.5) and (322.73,137.53) .. (322.73,135.1) -- cycle ;
\draw  [fill={rgb, 255:red, 255; green, 129; blue, 145 }  ,fill opacity=0.5] (355.73,136.9) .. controls (355.73,134.47) and (357.7,132.5) .. (360.13,132.5) -- (379.33,132.5) .. controls (381.76,132.5) and (383.73,134.47) .. (383.73,136.9) -- (383.73,150.1) .. controls (383.73,152.53) and (381.76,154.5) .. (379.33,154.5) -- (360.13,154.5) .. controls (357.7,154.5) and (355.73,152.53) .. (355.73,150.1) -- cycle ;
\draw [fill={rgb, 255:red, 255; green, 129; blue, 145 }  ,fill opacity=0.5]  (324.13,153.3) .. controls (324.13,150.87) and (326.1,148.9) .. (328.53,148.9) -- (347.73,148.9) .. controls (350.16,148.9) and (352.13,150.87) .. (352.13,153.3) -- (352.13,166.5) .. controls (352.13,168.93) and (350.16,170.9) .. (347.73,170.9) -- (328.53,170.9) .. controls (326.1,170.9) and (324.13,168.93) .. (324.13,166.5) -- cycle ;

\draw [line width=1.5pt]  (63.71,84.25) -- (133.71,84.25) -- (133.71,124.25) -- (63.71,124.25) -- cycle ;

\draw    (242,143.5) -- (315.62,66.45) ;
\draw [shift={(317,65)}, rotate = 493.69] [color={rgb, 255:red, 0; green, 0; blue, 0 }  ][line width=0.75]    (10.93,-3.29) .. controls (6.95,-1.4) and (3.31,-0.3) .. (0,0) .. controls (3.31,0.3) and (6.95,1.4) .. (10.93,3.29)   ;
\draw    (242,143.5) -- (316.33,143.5) ;
\draw [shift={(318.33,143.5)}, rotate = 180] [color={rgb, 255:red, 0; green, 0; blue, 0 }  ][line width=0.75]    (10.93,-3.29) .. controls (6.95,-1.4) and (3.31,-0.3) .. (0,0) .. controls (3.31,0.3) and (6.95,1.4) .. (10.93,3.29)   ;
\draw    (243.43,63) -- (316.97,142.04) ;
\draw [shift={(318.33,143.5)}, rotate = 227.06] [color={rgb, 255:red, 0; green, 0; blue, 0 }  ][line width=0.75]    (10.93,-3.29) .. controls (6.95,-1.4) and (3.31,-0.3) .. (0,0) .. controls (3.31,0.3) and (6.95,1.4) .. (10.93,3.29)   ;
\draw    (243.43,63) -- (316.33,63) ;
\draw [shift={(318.33,63)}, rotate = 180] [color={rgb, 255:red, 0; green, 0; blue, 0 }  ][line width=0.75]    (10.93,-3.29) .. controls (6.95,-1.4) and (3.31,-0.3) .. (0,0) .. controls (3.31,0.3) and (6.95,1.4) .. (10.93,3.29)   ;
\draw    (134,104.25) -- (199.08,142.49) ;
\draw [shift={(200.8,143.5)}, rotate = 210.44] [color={rgb, 255:red, 0; green, 0; blue, 0 }  ][line width=0.75]    (10.93,-3.29) .. controls (6.95,-1.4) and (3.31,-0.3) .. (0,0) .. controls (3.31,0.3) and (6.95,1.4) .. (10.93,3.29)   ;
\draw    (133.71,104.25) -- (198.3,64.06) ;
\draw [shift={(200,63)}, rotate = 508.11] [color={rgb, 255:red, 0; green, 0; blue, 0 }  ][line width=0.75]    (10.93,-3.29) .. controls (6.95,-1.4) and (3.31,-0.3) .. (0,0) .. controls (3.31,0.3) and (6.95,1.4) .. (10.93,3.29)   ;
\draw [line width=1.5pt]  (201,110.5) -- (243.43,110.5) -- (243.43,176.5) -- (201,176.5) -- cycle ;
\draw [fill={rgb, 255:red, 133; green, 186; blue, 248 }  ,fill opacity=0.5 ]  (207.97,122.61) .. controls (207.97,120.18) and (209.94,118.21) .. (212.37,118.21) -- (231.57,118.21) .. controls (234,118.21) and (235.97,120.18) .. (235.97,122.61) -- (235.97,135.81) .. controls (235.97,138.24) and (234,140.21) .. (231.57,140.21) -- (212.37,140.21) .. controls (209.94,140.21) and (207.97,138.24) .. (207.97,135.81) -- cycle ;
\draw [fill={rgb, 255:red, 133; green, 186; blue, 248 }  ,fill opacity=0.5 ]  (207.83,153.9) .. controls (207.83,151.47) and (209.8,149.5) .. (212.23,149.5) -- (231.43,149.5) .. controls (233.86,149.5) and (235.83,151.47) .. (235.83,153.9) -- (235.83,167.1) .. controls (235.83,169.53) and (233.86,171.5) .. (231.43,171.5) -- (212.23,171.5) .. controls (209.8,171.5) and (207.83,169.53) .. (207.83,167.1) -- cycle ;

\draw  [fill={rgb, 255:red, 133; green, 186; blue, 248 }  ,fill opacity=0.5 ] (63.93,210) .. controls (63.93,207.57) and (65.9,205.6) .. (68.33,205.6) -- (87.53,205.6) .. controls (89.96,205.6) and (91.93,207.57) .. (91.93,210) -- (91.93,223.2) .. controls (91.93,225.63) and (89.96,227.6) .. (87.53,227.6) -- (68.33,227.6) .. controls (65.9,227.6) and (63.93,225.63) .. (63.93,223.2) -- cycle ;
\draw  [fill={rgb, 255:red, 255; green, 129; blue, 145 }  ,fill opacity=0.5] (231.73,210.2) .. controls (231.73,207.77) and (233.7,205.8) .. (236.13,205.8) -- (255.33,205.8) .. controls (257.76,205.8) and (259.73,207.77) .. (259.73,210.2) -- (259.73,223.4) .. controls (259.73,225.83) and (257.76,227.8) .. (255.33,227.8) -- (236.13,227.8) .. controls (233.7,227.8) and (231.73,225.83) .. (231.73,223.4) -- cycle ;

\draw (180,10) node [anchor=north west][inner sep=0.75pt]   [align=left] {Edge cluster $\displaystyle C_{1}$};
\draw (180,180) node [anchor=north west][inner sep=0.75pt]   [align=left] {Edge cluster $\displaystyle C_{2}$};
\draw (312,10) node [anchor=north west][inner sep=0.75pt]   [align=left] {Data center $\displaystyle C_{3}$};
\draw (312,180) node [anchor=north west][inner sep=0.75pt]   [align=left] {Data center $\displaystyle C_{4}$};
\draw (60,65) node [anchor=north west][inner sep=0.75pt]   [align=left] {Gateway VM};
\draw    (70.71,96.25) .. controls (70.71,93.49) and (72.95,91.25) .. (75.71,91.25) -- (121.71,91.25) .. controls (124.48,91.25) and (126.71,93.49) .. (126.71,96.25) -- (126.71,111.25) .. controls (126.71,114.01) and (124.48,116.25) .. (121.71,116.25) -- (75.71,116.25) .. controls (72.95,116.25) and (70.71,114.01) .. (70.71,111.25) -- cycle  ;
\draw (75,97) node [anchor=north west][inner sep=0.75pt]   [align=left] {NGINX};
\draw (96.2,210) node [anchor=north west][inner sep=0.75pt]   [align=left] {Frontend microservice};
\draw (264,210) node [anchor=north west][inner sep=0.75pt]   [align=left] {Backend microservice};

\end{tikzpicture}}
    \caption{Emulated edge clusters and data centers hosting the two frontend and three backend microservices for the example federated face detection application.}
    \label{fig:edge_scenario}
    \noindent\makebox[\linewidth]{\rule{\linewidth}{0.4pt}}
\end{figure}

For the example application, frontend image pre-processing and backend classification where delivered in two standalone microservices. The frontend microservice reads a RGB image from an API call and reduces it to a grayscale image. The smaller grayscale image is sent to the backend microservice, where the face detection is performed using a standard Haar-cascade classifier in OpenCV. If a face is detected, the corners of the bounding box is returned to the frontend, which adds the box to the original RGB image and sends it back as a response to the original API call. The frontend microservices were then deployed with two replicas on each edge cluster, while the backend microservices used three replicas on each data center. Further, in order to balance requests between the multiple edge clusters, an NGINX load balancer was launched in a docker container on the Gateway VM. An overview of the scenario setup can be seen in Figure \ref{fig:edge_scenario}.

In order to put any real stress on the system, a load generator was further created which sends images at a given rate to the NGINX load balancer at the Gateway VM. For this demonstrative scenario, the necessary images where provided by the UMass face detection data set and benchmark\cite{Jain2010}.

\subsection{Experimental evaluation}

To see whether the system behaves as expected, a handful of experiments were performed on the example scenario. Initially, we examined how well TC-netem managed to create our desired network characteristics. This was done by measuring the RTT by running \texttt{ping} between every cluster simultaneously with an inter-sample time of 0.2~s over 600~s. We then examined if the cluster software stacks managed to properly handle the requests despite various disturbances, by observing the application vCPU usage at the different clusters when putting stress on the system. Thanks to the federation-wide monitoring, metrics such as vCPU for all clusters is easily observed in Grafana. 

The impact of two different inter-cluster load balancers in Istio, Round Robin (RR) and Least Connection (LC), under different loads was first examined. The RR load balancer distributes the requests evenly amongst the backends, which makes it largely unaffected by delay, while LC on the other hand prioritizes sending requests to the backend which has the least amount of active requests, making it sensitive to our added delay. Images were sent at both 10 images/s and 20 images/s for a time of 600 seconds per rate. If all is working properly, we would expect to see a vCPU usage lower for the edge clusters $C_1, C_2$ than for the data centers $C_3, C_4$ as these perform a more computationally heavy task. We further expect to see a doubling of vCPU usage when the load generation rate is doubled. Further, we expect to see roughly equal vCPU usage on $C_3, C_4$ using the RR policy, while using LC policy would prioritize $C_4$.

We then examined how well the emulation could reproduce sudden cluster link failure. The introduced delay was removed and RR load-balancing used for the inter-cluster load-balancing. The load generator was used to send 20 images/s to the system for 1200~s. At 300~s and 900~s the connection to cluster $C_3$ is suddenly lost for 300~s, by introducing a 100\% loss rate on the link of $C_3$ using TC-netem. We would in this case expect that the vCPU usage of $C_3$ reduce to 0, and the vCPU usage of $C_4$ to double during the outages.

\subsubsection{Results} 
\begin{figure*}[!t]
	\centering
	\subfloat[Using Istio's RR policy for inter-cluster load-balancing. \label{fig:grafana_data_rr}]{%
		\includegraphics[width=0.45\linewidth]{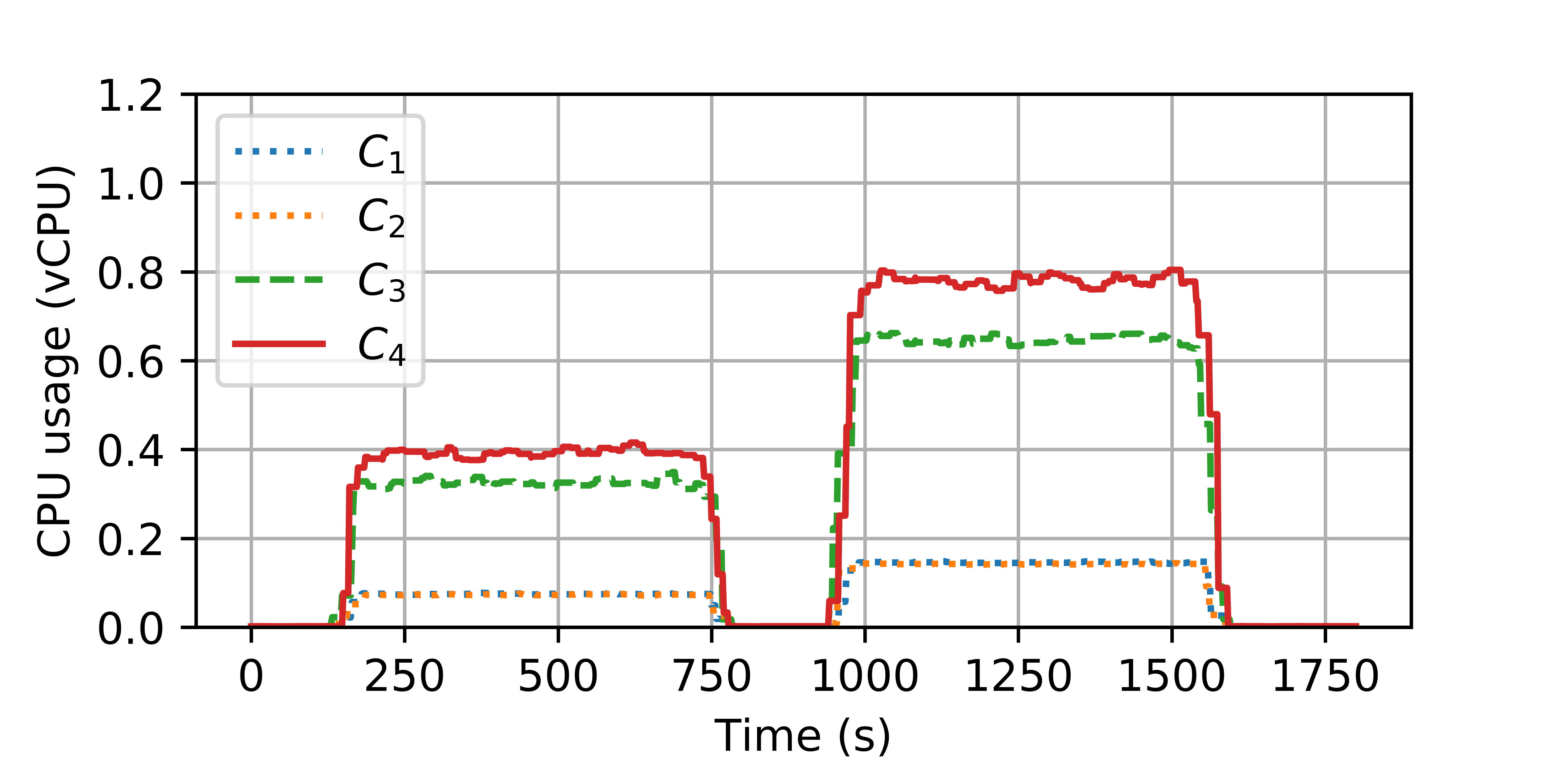}
    }
	\hspace{2em}
	\subfloat[Using Istio's LC policy for inter-cluster load-balancing. \label{fig:grafana_data_lc}]{%
		\includegraphics[width=0.45\linewidth]{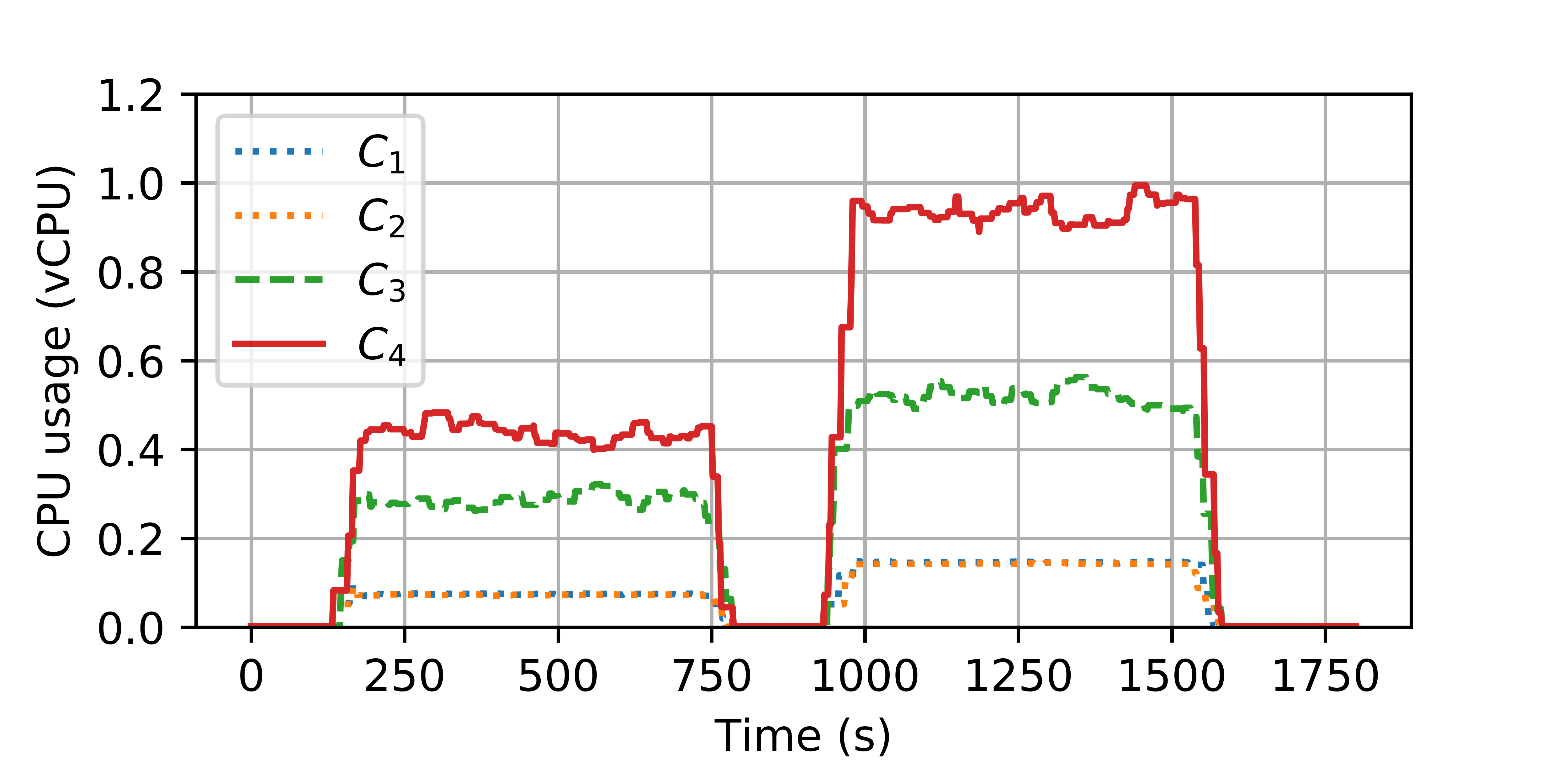}
    }
    \medskip\par
	\caption{CPU usage of the frontends on clusters 1 and 2, and the backends on clusters 3 and 4 under the two different load generation rates of 10 images/s and 20 images/s. The two graphs displays different inter-cluster load-balancing policies.}
	\label{fig:grafana_data_LB}
    \noindent\makebox[\linewidth]{\rule{\linewidth}{0.4pt}}
    \vspace{-2em}
\end{figure*}

\begin{figure*}[!t]
    \centering
    \includegraphics[width=\linewidth]{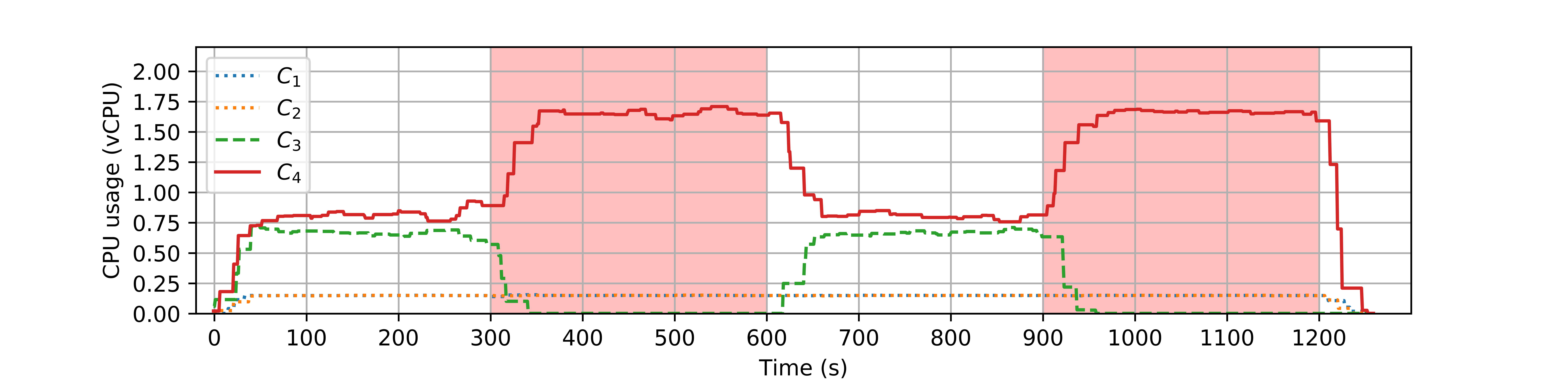}
    \caption{CPU usage of the frontends on clusters 1 and 2, and the backends on clusters 3 and 4 under a load generation rate of 20 images/s. Shaded red areas show when the connection to backend cluster 3 has been lost.}
    \label{fig:grafana_data_cfail}
    \noindent\makebox[\linewidth]{\rule{\linewidth}{0.4pt}}
\end{figure*}

For the RTT experiment, the resulting means and variances of the measured RTT for every link was
\begin{align*}
    \mathbb{E}\left[2\widehat{D}\right] \approx \begin{bmatrix}
        0.05 & 1.48 & 51.7 & 1.56 \\
        1.46 & 0.04 & 51.8 & 1.46 \\
        51.7 & 51.8 & 0.05 & 1.59 \\
        1.47 & 1.45 & 1.58 & 0.03
    \end{bmatrix}~ms
\end{align*}
As can be seen, the results corresponds well to the desired values. A small bias can be observed due to the internal OpenStack network. For our load generation experiments, the results can be seen in Figure \ref{fig:grafana_data_LB} and Figure \ref{fig:grafana_data_cfail}. For all graphs, the results looks as expected indicating that the sandbox works as intended. The slight difference in RR CPU usage could be contributed to the underlying VM allocations. As the measurement constitutes the rate of change in container CPU usage in seconds, a faster clock speed would lead to lower CPU usage. We can however be fairly certain that the RR difference is not due to the added delay, as the results in Figure \ref{fig:grafana_data_cfail} which has no delay experiences similar differences as the results displayed in Figure \ref{fig:grafana_data_rr}. Further, the apparent noisiness could be attributed to the different sizes of images in the dataset in conjunction with the inherent uncertainty in execution speed on the cloud. To let researchers try out this example by themselves, and to demo how a federation-wide application can be implemented and deployed in \sandbox{}, the example application is included in the sandbox.

\section{Related work} \label{sec:relatedwork}

In older literature, the matter of multi-cluster collaboration was focused on bridging technological gaps between competing cloud infrastructure providers with possibly heterogeneous underlying technology stacks~\cite{rochwerger2008reservoir}. The vision of inter-cloud compatibility was embodied by shared or compatible APIs on the infrastructure level~\cite{elmroth2009interfaces,edmonds2012toward,parak2014rocci}. With more than a decade of hindsight, it is clear that these early attempts failed to convincingly demand industry adoption among the major cloud providers. 

Recent work shows that a Kubernetes-based edge deployment is promising for real-life applications, and offers improvements with regard to e.g.\ latency~\cite{cicirelli2017edgebased,tsai2017distributed,huang2019design}. In addition to relying on Kubernetes to provide a common platform abstraction, cloud-native applications can leverage service meshes such as Istio to simplify networking code~\cite{sheikh2018modernize}. 

Compared to other approaches that aims to enable research for cloud systems, the sandbox differentiates itself in a number of ways. Its flexibility and openness puts it in stark contrast to other conceptually similar approaches, such as~\cite{kristiani2019implementation}. Further, in regards to the numerous Fog/Edge emulators, see Section 4.14.1 in \cite{Yousefpour2019}, cloud simulators such as CloudSim \cite{Calheiros2010}, or direct modeling of Kubernetes-based platforms, such as~\cite{medel2016adaptive,medel2018characterising}, the sandbox enables researchers to utilize the entire cluster software stack. We believe that due to the rapid pace of development in cloud orchestration tools such as Kubernetes and Istio, a practical sandbox that lets researchers deploy the actual software and not models thereof, have a greater impact and thus provide more valuable future insights.

\section{Discussion} \label{sec:discussion}

The way the emulated federation in our sandbox is constructed is motivated by our focus to support research in application orchestration. It is worth mentioning that our implementation choices are not the only way to create such an environment, especially with another focus in mind a different approach with different tools and topologies might be more suitable. For example, if one desires to test e.g. application scaling performance over a very large amount of clusters, the Gateway VM will be a potential issue.

This is because from a scalability perspective, the centralized connection constitutes a bottleneck. But scalability to support hundreds or more clusters is not part of our goal, and interesting problems to study as discussed in Section \ref{sec:researchfunc} does not need such large realizations. Moreover, the current implementation seems to have a maximum number of possible network interfaces for an OpenStack VM that would need to be circumvented for such a scenario. On our specific OpenStack instance we managed to allocate up to 22 clusters before running into this interface limit on our Gateway VM. A different OpenStack implementation or instance image might have a different interface limit. For these 22 clusters, we however experienced only negligible effects on network performance in a low traffic scenario, which hints that the sandbox can handle much larger experiments than our demonstrative scenario in Section \ref{sec:proofofconcept}. Ultimately, it would have been a nice feature to support an even larger amount of clusters but it is simply not the type of research that we have aimed to support in the current iteration of our work.

Besides our aims, there are other use-cases which the sandbox could support, or easily extended to support in its current form. For example, \sandbox{} has the potential to deliver an environment for both end-to-end and performance testing of specific multi-cluster software and tools under different scenarios. Kubernetes own federated deployment, KubeFed which is currently in its alpha comes to mind.

Looking forward, apart from our simple example it would be of benefit to out-of-the-box include other open source benchmark applications such as the ones introduced by Gan et al. \cite{Gan2019}. Further, implementing and deploying e.g. auto-scaling algorithms in Kubernetes is not a trivial matter for the inexperienced researcher. A benefit to usability would be to streamline this process as well, either via providing concrete examples or by some toolkit. 

\section{Conclusion}

We have presented \sandbox{}, a tool for providing an emulated cloud federation environment in OpenStack with user-defined network characteristics between clusters. The goal has been to give researchers a way to quickly set up federated environments for experimentation. Upon publication, the sandbox will be fully available as open source. 

\begin{acks}
This work is partially supported by the Wallenberg AI, Autonomous Systems and Software Program (WASP) funded by the Knut and Alice Wallenberg Foundation. Also, the work in this paper is partially supported by the SEC4FACTORY project, funded by the Swedish Foundation for Strategic Research (SSF), and the 5G-PERFECTA Celtic Next project funded by Sweden’s Innovation Agency (VINNOVA). Johan Ruuskanen \& Maria Kihl is part of the Excellence Center at Linköping-Lund on Information Technology (ELLIIT).
\end{acks}

\bibliographystyle{ACM-Reference-Format}
\bibliography{ms}

\end{document}